\documentstyle[12pt,epsf]{article}
\begin{document}

\font\fortssbx=cmssbx10 scaled \magstep2
\hbox to \hsize{
\includegraphics{/NextLibrary/TeX/tex/inputs/uwlogo.ps}
\hskip.5in \raise.1in\hbox{\fortssbx University of Wisconsin - Madison}
\hfill$\vcenter{\hbox{\bf MAD/PH/786}
            \hbox{August 1993}}$ }

\vspace{.25in}
\thispagestyle{empty}

\begin{center}
\large\bf 14th International Workshop on Weak Interactions: Concluding
Remarks\footnotemark
\\[.2in]
\normalsize Francis Halzen\\[.1in]
\it Department of Physics, University of Wisconsin, Madison, WI 53706
\end{center}

\footnotetext{Talk presented at the {\it 14th International Workshop on Weak
Interactions and Neutrinos}, Seoul, Korea, July 1993.}

\renewcommand{\LARGE}{\large}
\renewcommand{\Large}{\large}
\renewcommand{\thesection}{\arabic{section}.}

\vspace{.3in}

This superbly organized workshop invited the participants to focus on four
outstanding questions in weak interactions:
\begin{enumerate}
\item[i)] is the electroweak model correct at the quantum level?
\item[ii)] supersymmetry?
\item[iii)] neutrino mass?
\item[iv)] what is the nature of CP-violation?
\end{enumerate}
The meeting demonstrated how weak-interaction physics has become a terrain
successfully covered by accelerator and non-accelerator experiments in a very
complimentary way.

\section{Electroweak Theory}

A massive experimental effort is under way to test the Standard Electroweak
Model at the quantum level. High precision measurements of the properties of
the $Z$ at LEP and of the $W$ at $p\bar p$ colliders spearhead this effort.
Okun\cite{Okun} keeps reminding us that all electroweak observables are still
consistent with the ``Born-level" predictions provided these are evaluated
using the relevant coupling $\alpha(M_Z)\ (= 1/128.87(12))$ and not the low
energy Thomson charge $\alpha(0)\ (= 1/137.0359895(61))$. How does one obtain
precise determinations of the top mass (less than 20~GeV error in the analysis
presented in Ref.\cite{Hagiwara}; see also Refs.\cite{Brown,Kang,Park})
from 1~$\sigma$ measurements of the loop contributions to the measured
observables such as $\sin^2\theta_W$, asymmetries and the decay widths of the
$Z$-boson? The small 1$\sigma$ effects result from cancellation of large
positive contributions involving top with large negative corrections from all
other virtual particles in the loop: other quarks, weak bosons and the Higgs.
It is precisely the non-observation of electroweak corrections that places
stringent limits on the top mass.

This point should not be overemphasized. It is scheme-dependent, i.e.\ the
statements are only true in on-mass-shell renormalization. It is nevertheless
useful to measure quantities that depend on the top mass in an essentially
different way, i.e.\ not via purely oblique $t\bar t$ loop corrections. The
obvious observable is the decay width of the $Z$ into $b\bar b$. At this
conference the first measurements of this width were presented which achieved
a precision comparable to those of other LEP observables (Ref.\cite{Brown}).
Using $b$-tagging vertex detectors measurements of the forward-backward
asymmetry for $b$-quarks have also achieved a similarly high accuracy.
Exploiting polarization of the beams SLC has recently produced a measurement
of the Weinberg angle with a precision matching LEP.

This naturally raises the question of how far one can push the precision of the
determination of the top mass. Can one achieve sensitivity to the Higgs mass?
The significance of any determination of the Higgs mass with present statistics
is clearly illusionary. With steadily shrinking experimental errors it is
important, however, to remember that the theory has only made predictions at
the one-loop level. All the dominant two-loop effects have been evaluated and
computed but a complete calculation is unlikely. Two-loop effects are
important. For a given value of $M_W$ they shift the top mass by 5--20~GeV for
$m_t=120$--200~GeV. For a fixed value of $\sin^2\theta_W$ the mass shifts
associated with two-loop order are yet a factor 2 larger. If one is going to
make any claims on $m_t$ with precision 20~GeV, or on the value of the Higgs
mass, one must worry about the extend to which we know the two-loop radiative
corrections. The Achilles heel of the perturbative expansion is the threshold
contribution to the $t\bar t$ loop, symbolically shown below:

\begin{center}
\epsfxsize=5.5in\hspace{0in}\epsffile{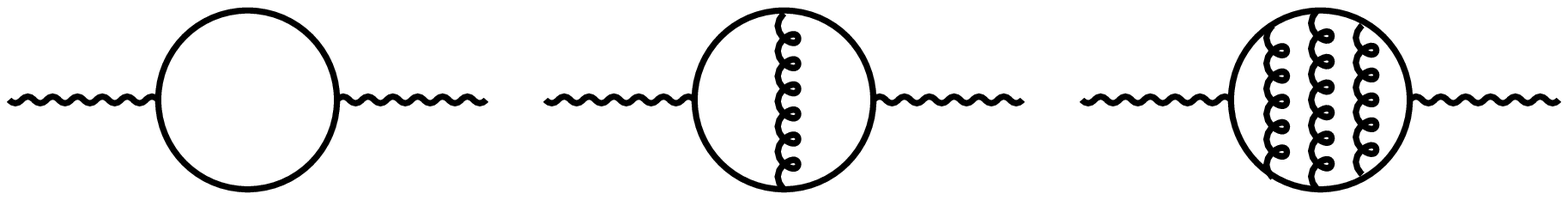}

{\small Fig.~1}
\end{center}

Although calculation to ${\cal O}(\alpha \alpha_s)$, i.e.\ the first two
diagrams is
straightforward, contributions from resonances below threshold and enhancement
of the cross section just above threshold are significant and their explicit
evaluation has turned out to be a problem. For the other quarks $q=u\dots b$
the problem is finessed by computing the loops directly from $e^+e^-\rightarrow
q\bar q$ data using dispersion relations. This procedure sums all orders of
perturbation theory. It, obviously, does not work for top. Although for heavy
tops the calculation of the threshold enhancements can be computed
perturbatively, such a computation is only possible in the non-relativistic
approximation. This turns out to be a problem when evaluating the threshold
contribution to the loop because of the slow convergence of the dispersion
relation\cite{Gonzalez}. Moreover, the required subtraction of the dispersion
relation is not unique\cite{Kniehl}. Two proposals in the literature do not
even yield the same sign of the top threshold contribution to the running of
$\alpha$. There is theoretical work to do in order to keep pace with
experiment!

\section{Supersymmetry?}

Once SU(5) predicted proton decay and it was not there. So, maybe there is no
unification, no desert and lots of physics at an energy scale just exceeding
those probed by our present accelerators. Strangely, theorists do not think
this way, they fix up the wrong prediction. It is clear what is
needed\cite{Hagiwara}:

\begin{enumerate}
\item[i)] introduce color exotics, e.g.\ leptoquarks, to extend the proton
lifetime;
\item[ii)] introduce multiple Higgses.
\end{enumerate}
Step i) cures rapid proton decay but also drives the Weinberg angle to
unacceptably low values and this can be fixed by introducing an extended Higgs
sector. Supersymmetry nicely provides both steps in a neatly packaged
form (Refs.\cite{Park,Ma}). The package also delivers:

\begin{enumerate}
\item[a)] a resolution of the bad behavior of radiative corrections in the
Standard Model. As for every boson there is a companion fermion, the bad
divergence associated with the Higgs loop shown below is cancelled by a fermion
loop with opposite sign,

\begin{center}
\epsfxsize=5.5in\hspace{0in}\epsffile{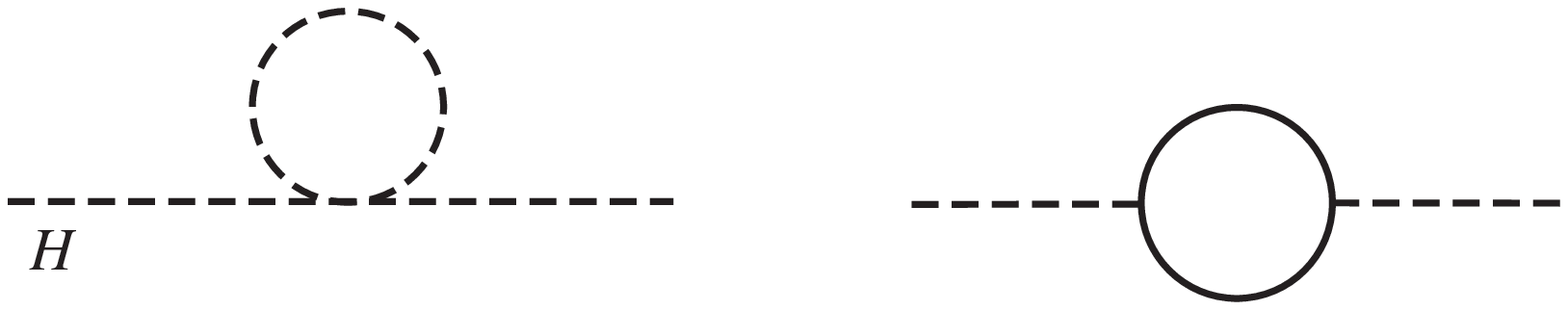}

{\small Fig.~2}
\end{center}

\item[b)] the lightest supersymmetric particle may be stable ($R$-parity) thus
providing us with an outstanding cold dark matter candidate
(Refs.\cite{JEKim,Krauss}).
\end{enumerate}

\noindent
Most importantly, in order to design future detectors we need dreams about
physics beyond the standard model. Supersymmetry is an intellectually pleasing,
yet specific and calculable dream. It takes some of the mystery out of the
puzzling separation of matter (half-integer spin) and forces (integer spin) in
the Standard Model. And, unlike the Standard Model, it can accommodate all
empirical constraints on the strong and electroweak couplings which nicely
unifies in the vicinity of $10^{16}$~GeV.

\section{Neutrino Mass}

Heroic efforts are pushing $\beta$-decay experiments towards their ultimate
sensitivity corresponding to an upper limit on the $\nu_e$ mass of a few eV
(Ref.\cite{Robertson}). The 17~KeV neutrino is no longer (Ref.\cite{Krauss}).
``Evidence" for neutrino mass is now based on the solar and atmospheric
``anomalies\rlap". The experimental evidence is, in both cases, conceptually
very similar. Underground experiments, sometimes the same experiment as in the
case of KAMIOKANDE, count a deficit of neutrinos from the sun and from the
atmospheric neutrino beam compared to the rate predicted by theoretical
calculations. The statistical weight of the evidence is very similar, with
several experiments supporting the existence of an anomalously low event count
in each case. I really do not see any reason to put these observations on a
different footing. The potential weakness of both is systematics. This is not
a criticism. Experimentalists are keenly aware of this fact and tackling the
problem, e.g.\ by the calibration of the KAMIOKANDE detector in a particle
beam. Another potential weakness is that the evidence, in both cases, is based
on the comparison of an experimental counting rate with a calculated one.
Although I have no special insights or any reasons to question the reliability
of the solar flux estimates, I do know as a particle physicist that one can
predict the ratio of electron-to-muon neutrinos in the atmospheric beam with
total confidence. Cosmic ray ambiguities drop out in the ratio which is
determined by textbook particle physics. So, here the atmospheric ``anomaly"
may be given a slight edge.

The opportunity exists, in principle, to eliminate the model dependence in
establishing the solar ``anomaly\rlap". The counting rate of solar neutrinos is
obtained by multiplying their solar flux with the detection cross section,
symbolically
\begin{equation}
N = \sum_i \sigma_i \phi_i \,.
\end{equation}
Here the sum is over the various cycles $i= pp,\ pep,\dots$Be, Bo. One can
convert this relation into an inequality which only involves the total flux of
the sun, which is unambiguously known\cite{Bahcall}
\begin{equation}
N < \sigma_{pp} \sum_i \phi_i = \sigma_{pp} \phi_\odot = 80\rm\ SNU \,.
\end{equation}

If an experiment like GALLEX or SAGE were to establish a flux below 80~SNU we
would be able to conclude that the solar ``anomaly" is a particle physics
problem without having to rely on modelling of solar fluxes. Unfortunately,
with the present values this is not the case, e.g.\ for GALLEX $N=97\pm23$
(Ref.\cite{Hahn}).

The problem with taking both anomalies seriously is that their interpretation
in terms of massive oscillating neutrinos predicts different mass values with
\begin{eqnarray}
&& 10^{-6}\mbox{--}10^{-10}\rm\,eV^2 \qquad \mbox{for the sun and},\\
&& 10^{-2}\mbox{--}10^{-3}\rm \,\,eV^2 \qquad\ \mbox{for the atmosphere}.
\end{eqnarray}
It is far from obvious how to incorporate both ``results" into a coherent
model (Ref.\cite{CWKim}). If you do not want to wait it out you have to be
pretty imaginative. Someone was (Ref.\cite{Halprin}).

\section{CP-violation and other weak interaction windows\hfil\break
 on high energy}

CP-problems come in the weak and strong variety
(Refs.\cite{Choi,Abe,Kobayashi}). In these talks and in a very lengthy and
exciting discussion session organized by C.~S.~Kim, the strategy was reviewed
in detail on how a $B$-factory can establish the fact that CP-violation is just
a consequence of the presence of a complex phase in the CKM matrix.
Confirmation of the Standard Model origin of CP-violation will make the
question loom very big why Nature supplied us with exactly the three
generations needed to introduce one complex phase allowing us to break CP and
make possible a ratio of baryon number to entropy of (4--6)${}\times 10^{-11}$
in the Universe, which is responsible for the fact that we are here pondering
the question. Alternatively, $B$-factories may disprove the Standard Model
origin of CP-violation, providing us with a new
window on physics beyond its reach.

With supercolliders embroiled in political turmoil and the technical
feasibility of linear electron-positron colliders unproven, it is good to
reassure ourselves that precision measurements in weak interactions have the
potential to indeed give us a glimpse of physics beyond the reach of
accelerators. Besides CP-violation, there is neutrino mass, double $\beta$
decay (Ref.\cite{Cremonesi}), $\mu \rightarrow e+\gamma$
experiments (Ref.\cite{Piilonen}). Even though evidence for Majorons or
Majorana neutrinos has been lacking, progress in double $\beta$ decay
experiments has moved them into a position to probe supersymmetry, for example,
with a sensitivity competitive with accelerators. On the theoretical side the
power of non-accelerator experiments to probe physics beyond the Standard
Model is pointedly illustrated by the intriguing result of the see-saw
model\cite{CWKim}
\begin{equation}
0.05{m_u^2\over m_I} : 0.08{m_c^2\over m_I} : 0.28{m_t^2\over m_I} =
m_{\nu_1} : m_{\nu_2} : m_{\nu_3} \,,
\end{equation}
which for $m_I=10^{13}$~GeV incorporates a 10~eV $\tau$-neutrino --- a
perfect candidate for the hot dark matter required by the COBE measurements of
the granularity of the cosmic photon background (Ref.\cite{Krauss}). The
coefficients in front of the mass ratios represent the running of the Yukawa
couplings. They were obtained by averaging the SU(5) and SO(10) predictions
which are, in fact, very similar.

\section{Return of the weak interactions and neutrinos as our window on high
energy}

One must conclude that the future of this field is bright. The variety of
experiments and the qualitative improvement in their sensitivity guarantee an
exciting transition period to the supercolliders. Even second generation
non-accelerator experiments will be commissioned for a price which is less
than 1\% of the SSC. Cheap experiments mean short timelines and data soon. At
the risk of omissions I close with a preview:

\begin{itemize}  

\item CP-violation:

\begin{itemize} 
\item $B$-factories,

\item $B$'s after Tevatron upgrade,

\item $\epsilon$ and $\epsilon'$ measurements.
\end{itemize}

\item Neutrino mass:

\begin{itemize}
\item oscillation experiments (CHORUS, NOMAD, new underground experiments using
the atmospheric beam),

\item long-baseline oscillation experiments (BNL and various
combinations of accelerator beams and deep underground detectors),

\item neutrino mass measurements, solar neutrino experiments (here we enter the
era of experiments with more that 10 events per day: SNO, BOREXINO,
ICARUS\dots;
observation of all neutrino types via neutral current detection),

\item high energy neutrino telescopes (AMANDA, BAIKAL, DUMAND and NESTOR),

\item supernova beams. Even in the latter case the timeline might be comparable
with the SSC if the error on their anticipated frequency in our galaxy is such
that one explodes every 25 years rather than once a century. A large variety of
much improved detectors will be ready (Ref.\cite{Sato}).
\end{itemize}

\item Other:

\begin{itemize}
\item the Universe,

\item edm moments,

\item rare decays with new $K$ and $\Phi$ factories,

\item double $\beta$ decay,

\item atomic parity violation, and

\item bolometric dark matter detectors.
\end{itemize}

\end{itemize}  

\noindent
A lot of exciting results will await us in Talloires in two years!

\section*{Acknowledgements}

I thank the organizers Professors Kang, Kim (C.S., J.E., S.K. and especially
Jewan) and Song for their excellent hospitality in the beautiful setting of
Seoul National University. I thank Manuel Drees and Concha Gonzalez for
comments on the manuscript.
This research was supported in part by the University of Wisconsin Research
Committee with funds granted by the Wisconsin Alumni Research Foundation, in
part by the U.S.~Department of Energy under contract DE-AC02-76ER00881 and in
part by the Texas National Research Laboratory Commission under Grant
No.~RGFY93-221.

\end{document}